\documentclass{elsart}
\usepackage{graphicx}
\usepackage{epsfig}

\begin{document}
\begin{frontmatter}
\title{Resultants in Genetic Linkage Analysis}
\author{Ingileif B. Hallgr\'{\i}msd\'ottir}
\address{Department of Statistics, University of California, Berkeley}
\author{Bernd Sturmfels}
\address{Department of Mathematics, University of California, Berkeley}

\begin{abstract}
Statistical models for genetic linkage analysis of $k$ locus diseases 
are $k$-dimensional  subvarieties of a $(3^k-1)$-dimensional 
probability simplex.  We determine the algebraic invariants of 
these models with general characteristics for $k=1$,  
in particular we recover, and generalize, the Hardy-Weinberg curve.
For $k = 2$, the algebraic invariants are presented as determinants of 
$32 \times 32$-matrices of linear forms in $9$ unknowns, a suitable 
format for computations with numerical data.
\end{abstract}
\end{frontmatter}

\section{Introduction}

Most common diseases have a genetic component.  The first step 
towards understanding a genetic disease is to identify the genes 
that play a role in the disease etiology. Genes are identified by their
location  within the genome.  \emph{Genetic linkage analysis}, or gene 
mapping \cite{ds,holmans,lander,ott},  
is concerned with this problem of finding the chromosomal 
location of  disease genes.  Over 1,200 disease genes
for have been successfully mapped~\cite{botrisch}, and this
has led to a much better understanding of
Mendelian (one gene) disorders. Most common diseases are, 
however, not caused by one gene but by  $k \geq 2$ genes. 
The challenge today is to understand complex diseases
(such as cancer, heart disease and diabetes) which are caused 
by many interacting genes and environmental factors.  

The human genome has approximately 25,000 genes.  Genes encode 
for proteins, and proteins perform all the cellular functions 
vital to life.  We all have the same set of genes, but there are 
many variants of each gene, called \emph{alleles}.  Usually these 
variants all produce a functional protein, but a mutation in a 
gene can change the protein product of the gene, and this may 
result in disease. Since mutations are rare, two affected 
siblings who have the same genetic disease probably inherited 
the same mutation from a parent.  Genetic linkage analysis makes 
use of this fact: one tries to locate disease genes by identifying 
regions in the genome that display statistically significant 
increased sharing across a sample of affected relatives, such as
sibling pairs~\cite{elston}.

The statistical models used in genetic linkage analysis are 
algebraic varieties. The given data are $k$-dimensional tables 
of format $3 \times 3 \times \cdots \times 3$. As usual in 
algebraic statistics (\cite{gss}, \cite{prw}, \cite[\S 7]{stbook}), 
there is one \emph{model coordinate} $z_{i_1 i_2 \cdots i_k}$ for each cell 
entry, where $i_1 ,i_2 ,\ldots,i_k \in \{0,1,2\}$.  This coordinate 
represents the probability that for an affected sibling pair
the IBD sharing (see section 2) at the first locus is $i_1$, the IBD 
sharing at the second locus is $i_2$, etc. 
The model is a subvariety of the probability simplex with 
these coordinates. It is $k$-dimensional, because the 
$z_{i_1 i_2 \cdots i_k}$ are given as  polynomials 
in $k$ \emph{model parameters} $\,p_1,p_2,\ldots,p_k$. 
Here $p_j$ represents the frequency  of the disease allele 
at the $j$-th locus. We consider an infinite family of 
models which depends polynomially on $3^k$  
\emph{model characteristics} $f_{i_1 i_2 \cdots i_k}$.  
The characteristic $\,f_{i_1 i_2 \cdots i_k}\,$ represents 
the probability that an individual who has $i_j$ copies 
of the disease gene at the $j$-th locus will get affected.  
Note  that the parameters $p_i$ and the characteristics 
$f_{i_1 i_2 \cdots i_k} $ 
are unknown, but we might be interested in estimating 
them from the given data~$z$.

This paper is organized as follows.  Section~\ref{oneloc} contains
a self-contained derivation of the models in the one-locus 
case $(k=1)$. Here the models are curves in a triangle with 
coordinates $(z_0,z_1,z_2)$. For general characteristics, $(f_0,f_1,f_2)$,
the curve has degree four. In Section~3 we compute 
its defining polynomial, a big expression in $z_0,z_1,z_2,f_0,f_1,
f_2$. This is done by elimination using the univariate
B\'ezout resultant.  We discuss what happens for special 
choices of characteristics
which have been studied in the genetics literature.

In Section~4 we derive the parametrization of 
the linkage models for $k \geq 2$.  In the two-locus 
case $(k=2)$, the models are surfaces in the space of 
nonnegative $3 \times 3$-tables $(z_{ij})$ whose entries 
sum to one. For general characteristics $(f_{ij})$,
the surface has degree $32$.  In Section~\ref{surf} 
we apply Chow forms to derive a system 
of \emph{algebraic invariants}.  These are 
the polynomials which cut out the surface.  Each 
invariant is presented  as the 
determinant of a $32 \times 32$-matrix whose 
entries are linear forms in the $z_{ij}$ whose 
coefficients depend on the $f_{ij}$.  We argue that 
this format is suitable  for statistical analysis 
with numerical data. Computational issues and further 
directions   are discussed in Section~6.

\section{Derivation of the One-Locus Model}
\label{oneloc}

\begin{figure}[h]
  \begin{center}
    \leavevmode
   \epsfig{file=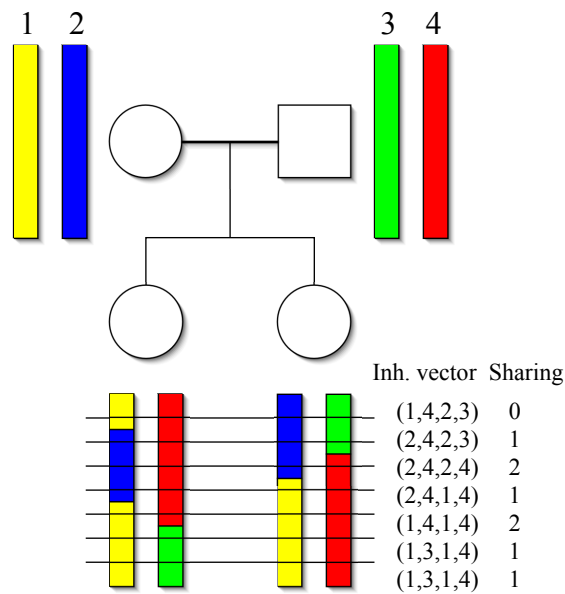, height=8cm}
    \caption{An example of the inheritance of one chromosome pair in parents and a sibling pair.   Squares represent males and circles females.}
    \label{fig:sibs}
  \end{center}
\end{figure}

The genetic code, the blueprint of life, is stored in our genome.  
The genome is arranged into chromosomes which can be thought of as 
linear arrays of genes.  The human genome has two copies of 
each chromosome, with 23 pairs of chromosomes,  22 autosomes 
and the sex chromosomes X and Y (women have XX and men XY). 
Each parent passes one copy of each chromosome to a child.
A chromosome passed from parent to child is a mosaic 
of the two copies of the parent, and a point at which the origin of a 
chromosome changes is called a \emph{recombination}.  This is illustrated in
Figure~\ref{fig:sibs}. 

Between any two recombination sites, the inheritance pattern
of the two siblings is constant and is encoded by
the {\em inheritance vector} $\,x=(x_{11}, x_{12}, 
x_{21}, x_{22})$. The entry $x_{kj}$ is the label of 
the chromosome segment that sibling $k$ got from parent $j$. 
If we label the paternal chromosomes with $1$ and $2$ and the 
maternal chromosomes with $3$ and $4$, then
$x_{11}, x_{21} \in \{1,2\}$ and $x_{12}, x_{22} \in \{3,4\}$, so 
there are 16 possible inheritance vectors $x$. 
They come in three classes:
\begin{eqnarray*}
C_0 \quad & = \quad &
 \bigl\{ \,
(1,3,2,4),\,
(1,4,2,3),\,
(2,3,1,4),\,(2,4,1,3)  \, \bigr\}, \\
C_1 \quad & = \quad &
 \bigl\{ \,
(1,3,1,4),\,
(1,4,1,3),\,
(2,3,2,4),\,
(2,4,2,3),\, \\ & & \,\,\,\,
(1,3,2,3),\,
(2,3,1,3),\,
(1,4,2,4),\,
(2,4,1,4)\, \bigr\} , \\
C_2 \quad &  = \quad &
 \bigl\{ \,
(1,3,1,3), \,
(1,4,1,4), \,
(2,3,2,3), \,
(2,4,2,4)\, \bigr\} .
\end{eqnarray*}

We say that two siblings share genetic material, at a locus,
identical by descent (IBD) if it originated from the same parent.
The IBD sharing at a locus can be 0, 1 or 2, where the inheritance
vectors in $C_i$ correspond to IBD sharing of $i$.  Since at a 
random locus in the genome each inheritance vector is equally likely 
the IBD sharing is 0, 1 or 2 with probabilities $1/4$, $1/2$ and $1/4$.

Each individual has two alleles, i.e. two copies of every gene, one on 
each chromosome.  A \emph{genotype} at a locus is the unordered pair 
of alleles.  We are 
only concerned with whether one carries an allele that predisposes 
to disease, which we call $d$, or a normal allele, called $n$.  
The set of possible genotypes at a disease locus is
 $\,G=\{nn, nd, dn, dd\}$.

Let $p$ denote the frequency of the disease allele 
$d$ in the population. This quantity is
our {\em model parameter}. We assume Hardy-Weinberg equilibrium:
$$ \hbox{$Pr(nn)=(1-p)^2, \, Pr(nd)=p(1-p), \, Pr(dn)=p(1-p)$ and $Pr(dd)=p^2$.}$$
A disease model is specified by
$f = (f_0,f_1,f_2)$, where $f_i$ is the probability 
that an individual is affected with the disease, given 
$i$ copies of the disease allele,
\vskip -0.3cm
\begin{eqnarray*}
f_0 &\,=\,& Pr(\mbox{affected} \,|\, nn), \quad f_2 \,=\, Pr(\mbox{affected} \,|\, dd), \\
f_1 &\,=\,& Pr(\mbox{affected} \,|\, nd ) \,= \, Pr(\mbox{affected} \,|\, dn).
\end{eqnarray*}
The quantities $f_i$ are known as {\em penetrances} in the
genetics literature. In this paper, we call them
{\em  model characteristics} to emphasize their algebraic role.

The {\em coordinates} of a disease model are $z = (z_0,z_1,z_2)$, where
 $z_i$ is the probability that the IBD sharing for an affected sibling pair 
is $i$ at a given locus, 
$$ z_i \,\, = \,\,  
Pr(\mbox{IBD sharing}=i \,|\, \mbox{both sibs affected}), \quad i=0,1,2. $$ 
Then, as was stated above, at a random locus not linked to 
the disease gene the distribution is $z_{null}=(1/4,1/2,1/4)$. 
Data for linkage analysis are collected from a sample of $n$ 
siblings (and parents) as follows.
The marker information is used to infer the IBD sharing at
each marker locus for each sibling pair and 
at any particular locus, one uses the
vector $(n_0,n_1,n_2)$, where $n_i$ is the number of
sibling pairs whose inferred IBD sharing is $i$ at the locus.
Each such data point determines an empirical distribution
$$ \hat{z} \,\, = \,\, (\hat{z}_0,\hat{z}_1,\hat{z}_2)  \,\, = \,\, (n_0/n,n_1/n,n_2/n) \, , \qquad \hbox{where}  \,\,\,\,
n_0+n_1+n_2 = n.  $$
The objective is to look for regions in the genome where $\,\hat{z}\,$
deviates significantly from  $\,z_{null} = (1/4, 1/2, 1/4)$.
Such regions may be linked to the disease.

The one-locus model is given by expressing the coordinates
 $(z_0,z_1,z_2)$ as polynomial functions of
the parameter $p$ and the characteristics $f_0,f_1,f_2$.
These polynomials are derived as follows. Consider
the set of events $\,\mathcal{E}_i \,=\, C_i \times G \times G\,$ for 
$i=0,1,2$.
Each event in $\mathcal{E}_i$ consists of
an inheritance vector, a genotype for 
the mother and a genotype for the father.
This triple determines the total number $m$
of disease alleles carried by the parents
and the numbers $k_1$ and $k_2$ of disease alleles
carried by the two siblings.
The probability of the event is
$$
f_{k_1} f_{k_2} p^m q^{4-m} \,, \quad \quad
\hbox{where $q = 1-p$.}
$$
Then, up to a global normalizing constant,
the IBD sharing probability $z_i$ is the sum over 
all events in $\mathcal{E}_i$ of the monomials
$\,f_{k_1} f_{k_2} p^m q^{4-m}$.
Hence $z_0$ is a sum of $|\mathcal{E}_0| = 64$ monomials,
$z_1$ is a sum of $ 128$ monomials,
and $z_2$ is a sum of $ 64$ monomials.
But these monomials are not all distinct. 
For instance, all four elements of
$\, C_0 \times \{nn\} \times \{nn\}\,\subset \,\mathcal{E}_0\,$ 
contribute the same monomial $\, f_0^2 q^4\,$ to $z_0$.
By explicitly listing all events in 
$\mathcal{E}_0, \mathcal{E}_1$ and $ \mathcal{E}_2$,
we get the following result.

\begin{prop} \label{matrixform}
The coordinates $z_i$ of the one-locus model
are homogeneous polynomials of bidegree 
$(2,4)$ in the characteristics
$(f_0,f_1,f_2)$ and the parameters $(p,q)$.
The column vector $(z_0,z_1,z_2)^T$ equals
the matrix-vector product
\vskip -.4cm
 \begin{eqnarray*}
\!\!\!\!\!\!
   \left( \begin{array}{ccccc}
4f_0^2 & 16f_0f_1 & 8f_0f_2+16f_1^2 & 16f_1f_2 & 4f_2^2 \\
8f_0^2 & 8(f_0^2 \!+\! 2f_0f_1 \!+\! f_1^2) & 
16 (f_0f_1\!+ \!f_1^2 \! + \! f_1f_2) & 
8(f_1^2\!+\!2f_1f_2\!+ \!f_2^2) & 8f_2^2 \\
4f_0^2 & 8f_0^2+8f_1^2 & 4f_0^2+16f_1^2+4f_2^2 & 8f_1^2+8f_2^2 & 4f_2^2
 \end{array} \right) 
\!\!
\left( \begin{array}{l}
q^4\\
pq^3\\
p^2q^2 \! \\
p^3q\\
p^4
\end{array} \right)
 \end{eqnarray*}
\end{prop}

Proposition \ref{matrixform} says that
the  one-locus model has the form
\begin{equation}
\label{zFq} (z_0,z_1,z_2)^T \,\, = \,\, F \cdot
(q^4, pq^3, p^2q^2, p^3q, p^4)^T , 
\end{equation}
where $F$ is a $3 \times 5$-matrix
whose entries are quadratic polynomials
in the penetrances $f_i$. The resultant
computation to be described in the
next section works for any model of this form, 
even if the matrix $F$ were more complicated.

\section{Curves in a Triangle}

Suppose that we fix the model characteristics
$f_0,f_1,f_2$ and hence the matrix $F$.
Then (\ref{zFq}) defines a curve in the projective
plane with coordinates $(z_0:z_1:z_2)$. The positive
part of the projective plane is identified with the
triangle
\begin{equation}
\label{bigtriangle}
 \bigl\{\,
(z_0,z_1,z_2) \,:\,
z_0,z_1,z_2 \geq 0 \,\,\, \hbox{and} \,\,\,
 z_0+z_1+z_2 = 1 \,\,\bigr\}. 
\end{equation}
The one-locus model with characteristics 
$f_0,f_1,f_2$ is the intersection of the curve
with the triangle. We are interested in its
defining polynomial.

\begin{prop} \label{twolocusprop}
For general characteristics $f_0,f_1,f_2$,
the one-locus model is a plane curve of degree four.
The defining polynomial of this
curve equals 
\vskip -.3cm
\begin{eqnarray*}
I(z_0,z_1,z_2) &\,\,=\,\,& 
 a_{1} z_0^3 z_2
+ a_{2} z_0^2 z_1^2
+ a_{3} z_0^2 z_1 z_2
+ a_{4} z_0^2 z_2^2
+ a_{5} z_0 z_1^3\\
& & 
+ \, a_{6} z_0 z_1^2 z_2
+ a_{7} z_0 z_1 z_2^2 
+ a_{8} z_0 z_2^3 
+ a_{9} z_1^4\\
& & 
+ \, a_{10} z_1^3 z_2
+ a_{11} z_1^2 z_2^2
+ a_{12} z_1 z_2^3
+ a_{13} z_2^4,
\end{eqnarray*}
where each $a_i$ is a polynomial  homogeneous
of degree eight in $(f_0,f_1,f_2)$.
\end{prop}

This proposition is proved by 
an explicit calculation. Namely,
the invariant $I(z_0,z_1,z_2)$ is gotten by
eliminating $p$ and $q$
from the three equations in (\ref{zFq}).
This is done using the \emph{B\'ezout resultant}
(\cite[Theorem 2.2]{StuSanDiego},
\cite[Theorem 4.3]{stbook}).
Specifically, we are using the
following $4 \times 4$-matrix from
\cite[Equation (1.5)]{StuSanDiego}:
\begin{equation}
\label{bezout}
\qquad \qquad B \,\,\, = \,\,\, 
 \left( \begin{array}{cccccccc}
& [12] & & [13]      & [14]      & & [15] & \\
& [13] & & [14] \! + \! [23] & [15] \! + \! [24] &  & [25] & \\
& [14] & & [15] \! + \! [24] & [25] \! + \! [34] & & [35] & \\
& [15] & & [25]      &  [35] & & [45] &
\end{array} \right).
\end{equation}

The determinant of this matrix is the {\em Chow form} \cite{DalStu} 
of the curve in projective $4$-space $P^4$ which is parameterized
by the vector of monomials $(q^4,pq^3,p^2 q^2, p^3 q, p^4)$.
We are interested in the curve in the projective plane $P^2$
which is the image of that monomial curve under the linear map from 
$P^4$ to $P^2$ given by the matrix $F$. Section 2.2 in  \cite{DalStu}
explains how to compute the image under a linear map of a variety 
that is presented by its Chow form. Applying the method described 
there means replacing the bracket $\, [i \, j]\,$ by
the $3 \times 3$-subdeterminant with column indices
$i$, $j$ and $6$ in the  matrix
from Proposition \ref{matrixform} augmented by $z$:
$$\! (F,z) =
   \left( \begin{array}{ccccccc}
\! 4f_0^2 & 16f_0f_1 & 8f_0f_2+16f_1^2 & 16f_1f_2 & 4f_2^2 && z_0 \\
\! 8f_0^2 & 8(f_0^2 \!+\! 2f_0f_1 \!+\! f_1^2) & 
16 (f_0f_1\!+ \!f_1^2 \! + \! f_1f_2) & 
8(f_1^2\!+\!2f_1f_2\!+ \!f_2^2) & 8f_2^2 & & z_1 \\
\! 4f_0^2 & 8f_0^2+8f_1^2 & 4f_0^2+16f_1^2+4f_2^2 & 8f_1^2+8f_2^2 & 4f_2^2
& & z_2 
 \end{array} \right) 
$$ 
The desired algebraic invariant equals
(up to a factor) the determinant of $\,B$:
\begin{equation}
\label{formulaforcurve}
  I(z_0,z_1,z_2)
 \,\, = \,\,
2^{-16} f_0^{-2} f_2^{-2} (f_0 - 2 f_1 + f_2)^{-4} \cdot {\rm det}(B).
\end{equation}

If the  characteristics $f_0,f_1,f_2$ 
are arbitrary real numbers between
$0$ and $1$ then the polynomial $\,I(z_0,z_1,z_2) \,$
is irreducible of degree four and its
zero set is precisely the model.
For some special choices of characteristics $f_i$, however,
the polynomial $I(z_0,z_1,z_2)$ may become reducible
or it may vanish identically. 
 In the reducible case,
the defining polynomial is one of the factors.
  Consider the following special 
models which are commonly used in genetics:
\begin{center}
\begin{tabular}{rcccc}
 & & $f_0$ &   $f_1$ &   $f_2$ \\
{\it dominant} &:&  0  & $f$ & $f$ \\[-2mm]
{\it additive} &:&  $0$ &  $f/2$ & $f$ \\[-2mm]
{\it recessive} &:& 0 & 0 & $f$ \\[-2mm]
\end{tabular}
\end{center}
Here $0 < f < 1$.  For the {\em dominant model}
our invariant specializes to
$$ I(z_0,z_1,z_2) \,\, = \,\,
4 f^8 (z_1-z_0-z_2)
(\underline{z_1^2 z_0-8 z_1 z_0 z_2
+4 z_1 z_2^2+4 z_0^2 z_2+4 z_0 z_2^2-4 z_2^3}),
$$
and the defining polynomial of the model is the underlined cubic factor.

For the {\em additive model}
our invariant specializes to
$$ 
I(z_0,z_1,z_2) \,\, = \,\,
\frac{f^8}{2^4}(z_1^2+2 z_1 z_2-8 z_0 z_2+z_2^2)
(\underline{z_1-z_0-z_2})^2 ,
$$
and the defining polynomial of the model is the underlined linear factor.

It can be shown that $\,I(z_0,z_1,z_2)\,$ vanishes identically if and only if
$$ f_0=f_1=0 \quad \hbox{or} \quad
f_1 = f_2 = 0 \quad \hbox{or} \quad
f_0=f_1 = f_2 . $$
This includes the {\em recessive model}, which is the familiar
Hardy-Weinberg curve:
$$z_1^2-4z_0z_2 \,\, = \,\, 0 .$$
\vspace{-0.5cm}
\begin{figure}[h]
  \begin{center}
    \leavevmode
    \epsfig{file=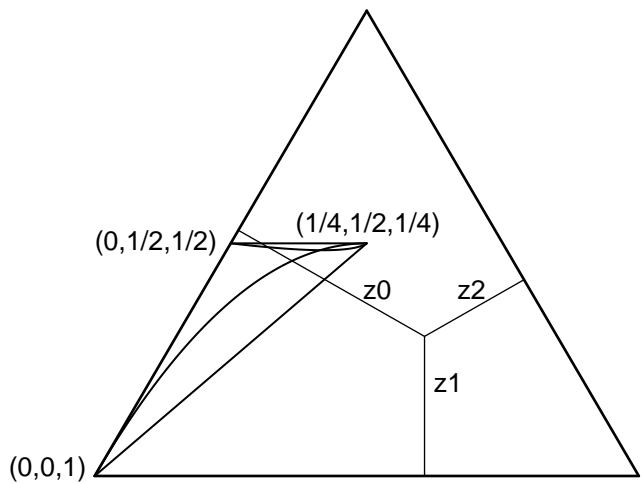, height=9cm, angle=270}
    \caption{Holmans' triangle.  The larger triangle is the probability simplex, $z_0+z_1+z_2=1$ and the smaller triangle is the possible triangle for sibling pair IBD sharing probabilities.  The curve from (1/4,1/2,1/4) to (0,0,1) is the Hardy-Weinberg (recessive) curve.  The curve from $(1/4,1/2,1/4)$ to $(0,1/2,1/2)$ is the dominant curve and the line between the same points is the additive curve.}
    \label{fig:holmans}
  \end{center}
\end{figure}
\newpage
Holmans \cite{holmans} showed that the IBD sharing probabilities 
for affected sibling pairs must satisfy 
$\,2 z_0 \leq  z_1 \leq z_0+z_2 $. This means we can restrict our 
attention to the  smaller triangle (Holmans' triangle) 
in Figure~\ref{fig:holmans}. 
 We can graph the curve in the triangle for any choice
of model characteristics.
 The part of the curve corresponding to values 
of $p \in [0,1]$ is within the smaller triangle.

\begin{figure}[h]
  \begin{center}
    \leavevmode
    \epsfig{file=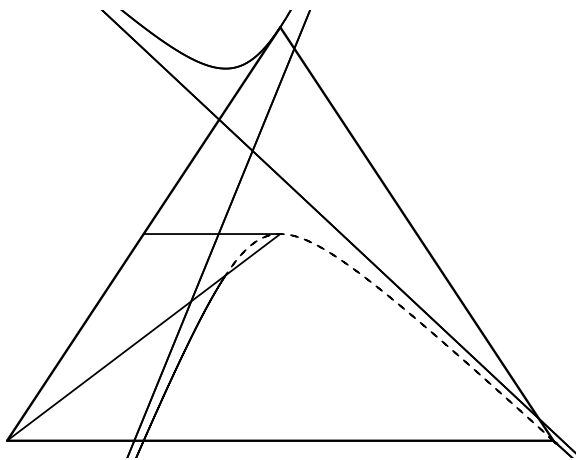, height=8cm, angle=270}
    \caption{Holmans' triangle.  The larger triangle is the probability simplex, $z_0+z_1+z_2=1$ and the smaller triangle is the possible triangle for sibling pair IBD sharing probabilities.  The curve corresponds to a model with characteristics $f_0=1, f_1=0$ and $f_2=1$.  The dotted part of the curve is the image of real valued $p$, and the solid part is the image of $\,p=1/2+y\sqrt{-1}$, for a real number $y$.}

    \label{fig:complex}
  \end{center}
\end{figure}

It is worth noting that not all points $(z_0,z_1,z_2)$ in Holmans' 
triangle which satisfy the algebraic invariant are in the image of
a point $(p,q)$ with real coordinates.
Consider e.g. the model with characteristics $f_0=1, f_1=0$ and $f_2=1$ 
and complex parameters $(p,q)$.  
The real part of the curve corresponding to this model is shown in 
Figure~\ref{fig:complex}.  Two segments of the curve are within 
Holmans' triangle, one of which (dotted) corresponds to values
$p \in [0,1]$.  The other segment has a complex pre-image.

\vspace{0.3cm}

We expressed the IBD sharing of the sibling pair at a gene locus 
(the model coordinate $z$) as a function of $f_0,f_1,f_2$ and $p$.
In practice, however, we get data at \emph{marker loci}, 
regularly spaced across the chromosomes, not at the gene locus.  
If there has been no recombination between the gene locus 
and a marker locus then the IBD sharing at the two loci is the same,
but different if there has been a recombination in either sibling.  
Let $\theta$ be the \emph{recombination fraction}
between the gene locus and the marker locus. The new parameter
$\theta$ depends on the distance between the two loci.  Following~\cite{ds},
we can express the IBD sharing probabilities at a marker locus 
distance $\theta$ away from the gene by the formula
\begin{equation}
\label{zFthetaq} (z_0,z_1,z_2)^T \,\, = \,\,  F_{\theta} \cdot
(q^4, pq^3, p^2q^2, p^3q, p^4)^T ,
\end{equation}
where $\,F_{\theta} = \Psi F \,$ and
\begin{eqnarray*}
\Psi \,\,\, = \,\,\, 
\left( \begin{array}{ccc}
\psi^2 & \bar{\psi} \psi & \bar{\psi}^2 \\
2 \bar{\psi} \psi & \psi^2 + \bar{\psi}^2 & 2 \psi \bar{\psi} \\
\bar{\psi}^2 & \bar{\psi} \psi & \psi^2
\end{array} \right), \quad
\hbox{with $\psi = \theta^2 + (1-\theta)^2$
and $\bar{\psi} = 1-\psi$.}
\end{eqnarray*}
One can easily repeat the resultant calculation in 
Proposition~\ref{twolocusprop} to obtain the equation of the larger family 
of curves defined by  (\ref{zFthetaq}).  Note that $\theta = 0$ corresponds 
to the earlier case, and increasing $\theta$  shifts the curve
towards $z_{null}$. 

We close this section with a statistical discussion. 
We wish to find the gene locus using the inferred IBD
sharing at the marker loci.  Since $\theta$ can be thought of 
as a measure of the distance between the marker locus and the
gene locus we wish to estimate $\theta$ at each marker locus.
The inferred IBD sharing can be used to obtain an estimate of the
model coordinates $z$.  If $p, f_0, f_1$ and $f_2$ are known it is 
then easy to estimate $\theta$.  However that is rarely the case, 
and it is impossible to identify all of the unknown quantities 
$p, f_0, f_1, f_2$ and $\theta$ from the coordinates $z$. 
Instead the model (\ref{zFq})
is applied to biological data as follows. 
The IBD sharing at the gene locus (and at nearby marker loci) 
is largest when the disease allele 
has a strong effect and/or the disease allele is rare, i.e. when
$f_0 \leq f_1 \leq f_2$ (and preferably $f_0 \ll f_2$),
and $p$ is small.  In these, biologically interesting,
situations the data point $\hat{z}$ is clearly different from $z_{null}$.
So in practice a test for genetic linkage tests whether $\hat{z}$ is
significantly different from $z_{null}$.  A widely used test statistic for 
linkage is $S_{pairs} = \hat{z}_2+\hat{z}_1/2$ which measures deviations 
from $z_{null}$ along the line corresponding to the additive model.

\section{Derivation of the Two-Locus Model}

Many common genetic disorders are caused by not one but many 
interacting genes.  We now consider the two-locus model, $k=2$, 
where we assume that two genes cause the disease, 
independently or together.  We shall assume that the genes are 
unlinked, i.e., they are either on different chromosomes or 
far apart on the same chromosome.  The derivation 
is much like in Section 2.

The {\em model parameters} are $p_1$ and $p_2$, where
$p_i$ is the  frequency of the disease allele at the $i$th locus.
 A two-locus genotype is an 
element in $G \times G = \{nn, nd, dn, dd\}^2$.
 The {\em model  characteristics} are
$\,f=(f_{00}, f_{01}, 
\ldots, f_{22})$ where $f_{ij}$,
is the probability that an individual is affected with the 
disease, given $i$ copies of the first disease allele and 
$j$ copies of the second disease allele:
\begin{eqnarray*}
f_{00} \,\, &= & \,\, Pr(\,\mbox{affected} \,\,\,|\,\,\, (nn, nn)\,), \\
f_{01} \,\, &= & \,\, Pr(\,\mbox{affected} \,\,\,|\,\,\, (nn, nd)\,)  \,\, = \,\, Pr(\,\mbox{affected} \,\,\,|\,\,\,(nn, dn)\,), \\
f_{02} \,\, &=& \,\, Pr(\,\mbox{affected} \,\,\,|\,\,\, (nn, dd)\,), \\
f_{10} \,\, &=& \,\, Pr(\,\mbox{affected} \,\,\,|\,\,\, (nd, nn)\,) \,\, = \,\, Pr(\,\mbox{affected} \,\,\,|\,\,\, (dn, nn)\,), \\
f_{11} \,\, &=& \,\, Pr(\,\mbox{affected} \,\,\,|\,\,\, (nd, nd)\,) \,\,= \dots = \,\,  Pr(\,\mbox{affected} \,\,\,|\,\,\, (dn, dn)\,), \\
f_{12} \,\, &=& \,\, Pr(\,\mbox{affected} \,\,\,|\,\,\, (nd, dd)\,) \,\, = \,\, Pr(\,\mbox{affected} \,\,\,|\,\,\, (dn, dd)\,), \\
f_{20} \,\, &=& \,\, Pr(\,\mbox{affected} \,\,\,|\,\,\, (dd, nn)\,), \\
f_{21} \,\, &=& \,\, Pr(\,\mbox{affected} \,\,\,|\,\,\, (dd, nd)\,) \,\,=\,\, Pr(\,\mbox{affected} \,\,\,|\,\,\, (dd, dn)\,), \\
f_{22} \,\, &=& \,\, Pr(\,\mbox{affected} \,\,\,|\,\,\, (dd, dd)\,).
\end{eqnarray*}
The {\em model coordinates} are 
$\,z=(z_{00}, z_{01}, z_{02}, z_{10}, z_{11}, z_{12}, z_{20}, z_{21}, z_{22})$,
where $z_{ij}$ represents the probability for an affected sibling pair
 that the IBD sharing at the first gene locus is $i$, 
 and $j$ at the second gene locus:
\begin{displaymath}
z_{ij} \,\, =  \,\, Pr(\,\mbox{IBD sharing}
\,\, = \,\, (i, j) \,|\, \mbox{both sibs affected}
\,), \qquad i,j = 0,1,2.
\end{displaymath}
The IBD sharing at two random loci, neither of which 
linked to the disease genes, is the null hypothesis 
$\,z_{null}~=~(1/16, 1/8, 1/16, 1/8, 1/4, 1/8, 1/16, 1/8, 1/16)$.

The polynomial functions which express the
coordinates $z_{ij}$ in terms of $p_1,p_2$ and the 
$f_{ij}$ are  derived as follows.
 We consider the set of events
$$ 
\mathcal{E}_i \times \mathcal{E}_j 
\,\, = \,\, 
C_i \times G \times G \times C_j \times G \times G
\quad \qquad \hbox{for $i,j=0,1,2$}. $$

Each event in $\,\mathcal{E}_i \times \mathcal{E}_j \,$ consists of an 
inheritance vector, the genotype of the father and the genotype 
of the mother, at each locus.  For a given event we know 
the total number $m_1$ and $m_2$ of disease alleles 
carried by the parents at the first and second locus 
and $k_{11}, k_{12}, k_{21}, k_{22}$, where 
$k_{ij}$ is the number of disease 
alleles carried by sibling $i$ at locus $j$.  The probability of the event is
\begin{displaymath}
f_{k_{11} k_{12}} f_{k_{21} k_{22}} p_1^{m_1}q_1^{4-m_1} p_2^{m_2} q_2^{4-m_2}, \quad \mbox{where} \quad q_1 = 1-p_1 \quad \mbox{and} \quad q_2 = 1-p_2. 
\end{displaymath}
Up to a normalizing constant,
 each IBD sharing probability $z_{ij}$ is the 
sum of the monomials $\,f_{k_{11} k_{12}} 
f_{k_{21} k_{22}} p_1^{m_1}q_1^{4-m_1} p_2^{m_2} q_2^{4-m_2}\,$ 
over all events in  $\,\mathcal{E}_i \times \mathcal{E}_j $.

\begin{prop} \label{matrixform2}
The coordinates $z_{ij}$ of the two-locus model
are homogeneous polynomials of tridegree 
$(2,4,4)$ in the characteristics $(f_0,f_1,f_2)$, 
the  parameters $(p_1,q_1)$ at the first locus, and
the  parameters $(p_2,q_2)$ at the second locus.
\end{prop}

The matrix form of the one-locus model given in 
Proposition \ref{matrixform} immediately generalizes
to the two-locus model. Let $\pi$ denote the
column vector whose entries are
the $25$ monomials of bidegree $(4,4)$ listed
in lexicographic order:
$$ \pi \,\, := \,\, \bigl(\,
 q_1^4 q_2^4,\,
 q_1^4 p_2 q_2^3,\,
 q_1^4 p_2^2 q_2^2,\,
\ldots\,,\,
p_1 q_1^3 q_2^4,\,
p_1 q_1^3 p_2 q_2^3,\,
\ldots, \,
p_1^4 p_2^4 \, \bigr).
$$

\begin{cor}  \label{ninetwentyfive}
The two-locus model has the form
$\,z^T = F \cdot \pi  \,$ where
$F$ is a $9\times 25$-matrix
whose entries are quadratic forms
in the characteristics $f_{ij}$.
\end{cor}

A typical entry in our $9 \times 25$ matrix $F$ looks like 
$$
32 \cdot ( f_{00}^2 + 2 f_{00} f_{10} + 4 f_{01}^2 
+ 8 f_{01} f_{11} +  f_{02}^2 + 2 f_{02} f_{12} 
+  f_{10}^2 + 4 f_{11}^2 +  f_{12}^2).  \eqno (*)
$$
This quadratic form appears in $F$ in row $6$ and column $8$.
It is the coefficient of the
$8^{th}$ biquartic monomial $\,p_1 q_1^3 p_2^2 q_2^2 \,$  in 
the expression for the $6^{th}$ coordinate:
\vskip -0.3cm
\begin{eqnarray*}
z_{12} &\quad=\,\,& \,\,\,\,(32 f_{00}^2) \cdot q_1^4 q_2^4  \,\,+\,\,
(64 f_{00}^2+64 f_{01}^2) \cdot q_1^4 p_2 q_2^3\\ 
& & + \, (32 f_{00}^2+128 f_{01}^2+32 f_{02}^2) \cdot q_1^4 p_2^2 q_2^2
\,+\, \cdots \cdots \, + \\
& & +\, (*) \cdot p_1 q_1^3 p_2^2 q_2^2 \,+\, \cdots\,
+ (64 f_{21}^2+64 f_{22}^2) \cdot p_1^4 q_2 p_2^3
\, +\, (32 f_{22}^2) \cdot p_1^4 p_2^4.
\end{eqnarray*}

\section{Surfaces of degree 32 in the 8-dimensional simplex}
\label{surf}

Let $\Delta_8$ denote the eight-dimensional probability simplex
$$
\{\,(z_{00},z_{01}, \ldots , z_{22}) \,\,
: \,\,
z_{ij} \geq 0 \,\, \mbox{for} \,\, i,j \in \{0,1,2\} 
\quad \mbox{and} \quad \sum_{i=0}^2 \sum_{j=0}^2 z_{ij} = 1\}.
$$
Likewise, we consider the
product of two $1$-simplices,
which is the square
$$ \Delta_1 \times \Delta_1 \,\,\, = \,\,\,
\bigl\{\, (p_1,q_1,p_2,q_2) \,\,:\,\,
p_1,q_1,p_2,q_1 \geq 0 \quad \mbox{and} \quad
p_1+q_1 = p_2 + q_2 = 1 \,\bigr\}. $$
For fixed $F$, 
the formula $\,z^T = F \cdot \pi  \,$  in
Corollary  \ref{ninetwentyfive}
specifies a polynomial map
$$ \tilde F \quad : \quad
\Delta_1 \times \Delta_1 \,\,\longrightarrow \,\,
\Delta_8 \qquad \qquad
\mbox{of bidegree $(4,4)$}. $$
The image of the map $\tilde F$
is the two-locus model 
for fixed characteristics $f_{ij}$.
The model is a surface in the simplex $\Delta_8$.
Our goal in this section is
to express this surface  as the common zero set
of a system of polynomials in the $z_{ij}$.

\begin{thm} \label{thirtytwo}
For almost all characteristics $f_{ij}$,
the two-locus model is a surface of degree
$32$ in the simplex $\Delta_8$. This surface is the
common zero set of the degree $32$ polynomials 
gotten by projection into three-dimensional subspaces.
\end{thm}

\noindent {\sl Proof. }
We work in the setting of complex projective
algebraic geometry. Consider the embedding
of the product of projective lines $P^1 \times P^1$
by the ample line bundle $\mathcal{O}(4,4)$. This
is a toric surface $X$ of degree $32$ in $P^{24}$.
The $9 \times 25$-matrix $F$ defines a rational 
map from $P^{24}$ to $P^8$, and it can be checked
computationally that this map has no base points on 
$X$ for general $f_{ij}$. Hence the image $F(X)$ of $X$ in 
$P^8$ is a rational surface of degree $32$. The two-locus model
is the intersection of $F(X)$ with $\Delta_8$, which is
the positive orthant in $P^8$.

Let $A$ denote a generic  $4 \times 9$-matrix,
defining a rational map $P^8 \rightarrow P^3$.
It has no base points on $F(X)$, hence the image
$AF(X)$ of $F(X)$ under $A$ is a surface
of degree $32$ in projective $3$-space $P^3$.
The inverse image of $AF(X)$ in $P^8$
is an irreducible hypersurface of degree
$32$ in $P^8$. It is defined
by an irreducible homogeneous polynomial
of degree $32$ in $\, z = (z_{00}, z_{01}, \ldots,z_{22})$.
These polynomials for various $4 \times 9$-matrices $A$
are known as the \emph{Chow equations} of the surface $F(X)$.
Computing them is equivalent to computing the 
\emph{Chow form} of $F(X)$. A well-known
construction in algebraic geometry (see e.g.~\cite[\S 3.3]{DalStu})
shows that any irreducible projective variety
is set-theoretically defined by its
Chow equations. Applying this result
to $F(X)$ completes the proof. \qed

We now explain how Theorem \ref{thirtytwo}
translates into an explicit algorithm for 
computing the algebraic invariants of the
two-locus model. Let $\,\mathcal{R}_X\,$ be
the Chow form of the toric surface
$\, X \simeq P^1 \times P^1 \,$ in $\,P^{24}$.
The Chow form $\,\mathcal{R}_X\,$ 
is the multigraded resultant of three polynomial equations
of bidegree $(4,4)$:
$$
\sum_{i=0}^4 \sum_{j=0}^4 \alpha_{ij} x^i y^j
\,=\,
\sum_{i=0}^4 \sum_{j=0}^4 \beta_{ij}  x^i y^j
\,=\,
\sum_{i=0}^4 \sum_{j=0}^4 \gamma_{ij}  x^i y^j
\,=\, 0 . $$
In concrete terms,  $\,\mathcal{R}_X\,$ is 
the unique (up to sign) irreducible polynomial
of tridegree $(32,32,32)$ in the
$75$ unknowns $\alpha, \beta,\gamma$ which vanishes
if and only if the three equations have a common 
solution in $\,P^1 \times P^1$. 

We use the B\'ezout matrix representation 
of the resultant $\mathcal{R}_X$
given in \cite[Theorem 6.2]{DicEmi}.
This is a $32 \times 32$-matrix  ${\bf B}$ which is 
a direct generalization of the $4 \times 4$-matrix
in (\ref{bezout}). Consider the
$3 \times 25$-coefficient matrix
$$ 
   \left( \begin{array}{cccccccccc}
\alpha_{00} & \alpha_{01} & \alpha_{02} & \alpha_{03} & \alpha_{04} &
\alpha_{10} & \alpha_{11} & \cdots \cdots & \alpha_{43} & \alpha_{44} \\
\beta_{00} & \beta_{01} & \beta_{02} & \beta_{03} & \beta_{04} &
\beta_{10} & \beta_{11} & \cdots \cdots & \beta_{43} & \beta_{44} \\
\gamma_{00} & \gamma_{01} & \gamma_{02} & \gamma_{03} & \gamma_{04} &
\gamma_{10} & \gamma_{11} & \cdots \cdots & \gamma_{43} & \gamma_{44} \\
\end{array} \right)
$$
For $1 \leq i < j < k \leq 25$, let $\,[\, i \,j \, k \,]\,$ denote the
determinant of the $3 \times 3$-submatrix with column indices $i,j,k$.
The entries in the Bezout matrix ${\bf B}$
are the linear forms in the brackets
$\,[\, i \,j \, k \,]$, and we have
$\,\mathcal{R}_X = {\rm det}({\bf B})$.

Let $F$ be the $9  \times 25$-matrix 
in Corollary \ref{ninetwentyfive}.
We add the column vector $z$ to get the
$ 9 \times 26$-matrix $\,( F \, z )$.
Next we pick any $4 \times 9$-matrix $A$
and we consider 
$$ A \cdot (F \,\, z) \,\, = \,\, (A \cdot F \, \,\,\, A \cdot z). $$
This is a $4 \times 26$-matrix whose last column consists of 
linear forms in the $z_{ij}$. 

In the B\'ezout matrix ${\bf B}$, we now replace 
each bracket $\,[\, i \,j \, k \,]\,$ by the $4 \times 4$-subdeterminant
of $\, A \cdot (F \, \, z)\,$ with column indices 
$i,j,k$ and $26$. Thus  $\,[\, i \,j \, k \,]\,$ is a linear
form in the $z_{ij}$ whose coefficients are homogeneous
polynomials of degree six in the $f_{ij}$. 
The matrix gotten by this substitution is
denoted $\,{\bf B}\bigl(A \cdot (F \,\, z) \bigr)$.
Its determinant is the specialized resultant
$\,\mathcal{R}_X \bigl( A \cdot (F \,\, z) \bigr)$.

\begin{cor}
The resultant $\,\mathcal{R}_X \bigl( A \cdot (F \,\, z) \bigr)\,$
is a homogeneous polynomial of degree $32$ in the entries $a_{ij}$ of $A$.
Its coefficients are polynomials which are bihomogeneous of degree $32$
in the $z_{ij}$ and degree $192$ in the $f_{ij}$.
The two-locus model is cut out by this finite list of coefficient polynomials
in the $z_{ij}$ and $f_{ij}$. \end{cor}

\noindent {\sl Proof. }
Each entry of the $32 \times 32$-matrix 
$\,{\bf B}\bigl(A \cdot (F \,\, z) \bigr)\,$ is
a polynomial which is trihomogeneous of degree
$(1,6,1)$ in $(a_{ij},f_{ij},z_{ij})$. Hence its determinant
is trihomogeneous of degree $(32,192,32)$.
For fixed $A$ and fixed $F$, the resulting polynomial
defines a hypersurface of degree $32$ in $P^{24}$.
This hypersurface is the inverse image of the
surface $AF(X)$ in $P^3$. As discussed in the 
proof of Theorem \ref{thirtytwo}, our model is
the intersection of these hypersurfaces
for all possible choices of $A$. A finite basis for
the linear system of these hypersurfaces
is given by the coefficient polynomials
of $\,\mathcal{R}_X \bigl( A \cdot (F \,\, z) \bigr)\,$
with respect to $A$. \qed

The finite list of algebraic invariants described in the
previous corollary is the two-locus generalization 
of the one-locus invariant in Proposition
\ref{twolocusprop}.
Note that the bidegree in $(F,z)$ has now
increased from $(4,8)$ to $(32,192)$.
Our derivation of these invariants
from the Chow form of a Segre-Veronese variety
generalizes to the $k$-locus case,
where $F$ and $z$ are $k$-dimensional tables of format $3 \times 3 \times \cdots \times 3$.
The analogous invariants have bidegree
$\,\bigl( \,k ! \, 4^k,\, 2 (k+1)! \, 4^k \,\bigr) \,$ in $(z,F)$.

\section{Computational experiments and statistical perspectives}
We prepared a test implementation in {\tt maple} of the elimination 
technique described in the previous section. That code is available
at the first author's website {\tt www.stat.berkeley.edu/$\sim$ingileif/}.
The input is a triple 
$\bigl((f_{ij}), (z_{ij}),A\bigr) $ consisting of
a $3 \times 3$-matrix of model characteristics,
a $3 \times 3$-matrix of model coordinates.
and a projection matrix of size $4 \times 9$.
Each entry in these input matrices can be either 
left symbolic or it can be specialized to a number.
Our program builds the specialized B\'ezout matrix 
$\,{\bf B}\bigl(A \cdot (F \,\, z) \bigr)$, and, if the
matrix entries are purely numeric, then 
it evaluates  the determinant $\,\mathcal{R}_X \bigl( A \cdot (F \,\, z) \bigr)$.

Here are some examples of typical
computations with our {\tt maple} program.
Set  \vskip -0.4cm
\begin{tabbing}
$\quad$ \= $z_{00} = 3 \quad$ \= $z_{01} = 3 \quad$ \= $z_{02} = 5 \quad$  \= $\quad$ \= $f_{00} = 32 \quad$ \= $f_{01} = 21 \quad$ \= $f_{02} = 48 \quad$ \\
\> $z_{10} = 29$ \> $z_{11} = 11$ \> $z_{12} = 13$ \> $\quad$ \> $f_{10} = 14$ \> $f_{11} = 27$ \> $f_{12} = 39$ \\
\> $z_{20} = 17$ \> $z_{21} = 19$ \> $z_{22} = 23$ \> $\quad$ \> $f_{20} = 36$ \> $f_{21} = 19$ \> $f_{22} = 22$ \\
\end{tabbing}
\vskip -0.4cm
$$ \hbox{and}
\qquad \qquad A \,\,\, = \,\,\,
 \left( \begin{array}{cccccccccc}
 1 & 0 & 0 &  0 & 0 & 0 &  0 & 0 & 0 \\
  0 & 1 & 0 &  0 & 0 & 0 &  0 & 0 & 0 \\
  0 & 0 & 0 &  1 & 0 & 0 &  0 & 0 & 0 \\
  0 & 0 & 0 &  0 & 1 & 0 &  0 & 0 & 0 \\
    \end{array} \right). \qquad \qquad \qquad \qquad $$
Then $\,{\bf B}\bigl(A \cdot (F \,\, z) \bigr)$ is a $32 \times 32$-matrix whose
entries $b_{i,j}$ are integers, e.g.,
$$ b_{1,1} =  26967093018624, \,\,b_{1,2} =  -114552012275712, \ldots, \,\,b_{32,32} =  845647773696. $$
The determinant of this $32 \times 32$-matrix is a non-zero integer with $469$ digits:
$$ \mathcal{R}_X \bigl( A \cdot (F \,\, z) \bigr) \,\, = \,\,                                    
                                                    0.2704985126... \cdot 10^{469}. $$
We now retain the numerical values for the model characteristics $f_{ij}$
and the  matrix $A$ from before but we make the model
coordinates $z_{ij}$ indeterminates. Then
$\,{\bf B}\bigl(A \cdot (F \,\, z) \bigr)$ is a $32 \times 32$-matrix whose
entries $b_{i,j}$ are linear forms
\vskip -0.4cm 
\begin{eqnarray*}
b_{1,1} &\,\,  = \,\, &  -2630935904256 \, z_{00}+1315467952128 \, z_{01} \\
 & &  +1315467952128 \, z_{10}-657733976064 \, z_{11} \\
b_{1,2} &\,\, = \,\,&  11746198683648 \, z_{00}-8211709034496 \, z_{01} \\
&  &  -5873099341824 \, z_{10}+4105854517248 \, z_{11}\\
&  & \qquad \dots  \quad \dots \quad \dots  \quad \dots \quad \dots 
\end{eqnarray*} \vskip -0.3cm
Its determinant  $\mathcal{R}_X \bigl( A \cdot (F \,\, z) \bigr)$ is an irreducible
polynomial of degree $32$
which vanishes on the model with the given characteristics $f_{ij}$.
In fact, up to scaling, it is the unique such polynomial 
which depends only on $\,z_{00},z_{01},z_{10}$ and $z_{11}$.

Finally, we reverse the role of the coordinates $z_{ij}$
and the characteristics $f_{ij}$, namely, we fix the former
at their previous numerical values $(z_{00} =3,\ldots,z_{22} = 22)$
but we regard the $f_{ij}$ as indeterminates. Then $\,{\bf B}\bigl(A \cdot (F \,\, z) \bigr)$
 is a $32 \times 32$-matrix whose entries $b_{i,j}$ are
 homogeneous polynomials of degree six, e.g.,
 \vskip -0.3cm 
\begin{eqnarray*}
 b_{1,1} &\,\, =\,\, & \quad 671744 \, f_{00}^6-1343488 \, f_{00}^5 f_{01}-1343488 \, f_{00}^5 f_{10}\\
        &       & + \, 671744 \, f_{00}^4 f_{01}^2  + 2686976 \, f_{00}^4 f_{01} f_{10}+671744 \, f_{00}^4 f_{10}^2 \\
        &       & - \, 1343488 \, f_{00}^3 f_{01}^2 f_{10}-1343488 \, f_{00}^3 f_{01} f_{10}^2 + 671744 \,f_{00}^2 f_{01}^2 f_{10}^2.
\end{eqnarray*}
Now  $\mathcal{R}_X \bigl( A \cdot (F \,\, z) \bigr)$ is an irreducible homogeneous
polynomial of degree $192$ in the nine characteristics $f_{ij}$.
The vanishing of this polynomial provides an algebraic constraint on the
set of all models $(f_{ij})$ which fit the given data $(z_{ij})$.

In linkage analysis, the characteristics $f_{ij}$ can take on any 
real value between $0$ and $1$. 
Two-locus models are often constructed by 
first picking two one-locus  characteristics, $g=(g_0, g_1, g_2)$ and $h=(h_0, h_1, h_2)$, from a class of special models such as recessive or dominant.
Then the two-locus model is defined by combining the one-locus characteristics in one of the following ways:
\begin{center}
\begin{tabular}{rcl}
{\it multiplicative} &:& $f_{ij} \,=\, g_i \cdot h_j$ \\
{\it heterogeneous} &:& $f_{ij} \,=\, g_i + h_j -g_i\cdot h_j$ \\
{\it additive} &:& $f_{ij} \,=\, g_i + h_j$ \\
\end{tabular}
\end{center}
The $9 \times 25$-matrix $F$ of the multiplicative model
is the tensor product of the two $3 \times 5$-matrices gotten
 from $g$ and $h$ as in Proposition \ref{matrixform}.
Hence the surface of the multiplicative  model is the
\emph{Segre product} of two one-locus curves.
The heterogeneous model and the additive model are too special,
in the sense that the corresponding surfaces in $P^8$ have degree
less than $32$. In these two cases, the resultant
$\,\mathcal{R}_X \bigl( A \cdot (F \,\, z) \bigr)\,$ vanishes
identically, and our {\tt maple} code always outputs zero.
The surfaces arising from these two models require
a separate algebraic study. Conducting this study could be
a worthwhile next step.

The following two-locus analogue to Holmans' triangle (the smaller triangle
in Figure \ref{fig:holmans}) was derived in~\cite{olof}.  For affected sibling pairs the IBD sharing probabilities $ \,z = (z_{00}, z_{01}, \ldots, z_{22})\,$ 
 satisfy
$\, H \cdot z^T \geq 0 \,$ where $H$ is the inverse of $K^{\otimes 2}$ and
\vskip -0.5 cm 
\begin{eqnarray*}
K &\,\,\, = \,\,\, & \frac{1}{4}
\left( \begin{array}{rrr}
 1 & 0 & 0 \\
 2 & 2 & 0 \\
 1 & 2 & 4 \\
\end{array} \right) 
\end{eqnarray*}
\vskip -0.5 cm
So, in practical applications we are only interested in the
intersection of our degree $32$ surface with the $8$-simplex defined by
these linear inequalities.

In summary, in this paper we have presented a model for the sharing 
of genetic material of two affected siblings, used in genetic linkage 
analysis, in the framework of algebraic geometry.  
The model is rich in structure, but this 
structure is not yet fully exploited in statistical tests for genetic linkage.  
For plausible biological models we expect to see increased sharing between 
affected sibling pairs at gene loci linked to the disease.  
The null hypothesis for linkage is rejected only if the estimate 
of the model coordinates, $z$, differs significantly from $z_{null}$. 
This is a geometric statement about the
distance between two points in a triangle (for $k=1$) or
in an $8$-simplex (for $k=2$). We believe that the algebraic
representation of the model derived here will be useful for 
deriving new test statistics for linkage in the case when $k \geq 2$.

\section{Acknowledgements}
We thank Lior Pachter and Terry Speed for 
reading the manuscript and providing useful 
comments. We are grateful to Amit Khetan
for helping us with the {\tt maple} implementation
of the B\'ezout resultant.  Bernd Sturmfels was supported
by the Hewlett Packard Visiting Research Professorship 2003-04
at MSRI~Berkeley and
the National Science Foundation (DMS-0200729).

\end{document}